\documentclass[conference]{IEEEtran}
\IEEEoverridecommandlockouts

\usepackage{cite}
\usepackage{amsmath,amssymb,amsfonts}
\usepackage{algorithmic}
\usepackage{graphicx}
\usepackage{textcomp}
\usepackage{xcolor}
\def\BibTeX{{\rm B\kern-.05em{\sc i\kern-.025em b}\kern-.08em
    T\kern-.1667em\lower.7ex\hbox{E}\kern-.125emX}}
\begin{document}

\title{SpikeWFM: Spiking-Aided Wireless Foundation Model for  Robust Channel Prediction

\thanks{This work is supported by Mobile Information Networks-National Science and Technology Major Project (Project No. 2025ZD1304900).	
\textsuperscript{\textdagger} These authors contributed equally: Liwen Jing and Yisha Lu. 
\textsuperscript{*} Correspondence to: Tingting Yang.}
}

\author{\IEEEauthorblockN{1\textsuperscript{st} Liwen Jing\textsuperscript{\textdagger}}
\IEEEauthorblockA{
	\textit{Pengcheng Laboratory}\\
	Shenzhen, China \\
	jinglw@pcl.ac.cn}
\and
\IEEEauthorblockN{2\textsuperscript{nd} Yisha Lu\textsuperscript{\textdagger}}
\IEEEauthorblockA{
\textit{Pengcheng Laboratory}\\
Shenzhen, China \\
luyisha@sztu.edu.cn}
\and
\IEEEauthorblockN{3\textsuperscript{rd} Tingting Yang\textsuperscript{*}}
\IEEEauthorblockA{
	\textit{Pengcheng Laboratory}\\
	Shenzhen, China \\
	yangtt@pcl.ac.cn}
\and
\IEEEauthorblockN{4\textsuperscript{th} Li Sun}
\IEEEauthorblockA{
	\textit{Pengcheng Laboratory}\\
	Shenzhen, China \\
	sunl03@pcl.ac.cn}
\and
\IEEEauthorblockN{5\textsuperscript{th} Yuxuan Shi}
\IEEEauthorblockA{
	\textit{Pengcheng Laboratory}\\
	Shenzhen, China \\
	shiyx01@pcl.ac.cn}
\and
\IEEEauthorblockN{6\textsuperscript{th} Yuwei Wang}
\IEEEauthorblockA{
	\textit{Pengcheng Laboratory}\\
	Shenzhen, China \\
	wangyw03@pcl.ac.cn}

\and
\IEEEauthorblockN{7\textsuperscript{th} Mengfan Zheng}
\IEEEauthorblockA{
	\textit{Pengcheng Laboratory}\\
	Shenzhen, China \\
	zhengmf01@pcl.ac.cn}

\and
\IEEEauthorblockN{8\textsuperscript{th} Leiyang Xu}
\IEEEauthorblockA{
	\textit{Pengcheng Laboratory}\\
	Shenzhen, China \\
	xuly@pcl.ac.cn}
}
\maketitle

\begin{abstract}
This paper proposes SpikeWFM, a novel hybrid architecture that integrates spiking neural networks (SNNs) with conventional artificial neural network (ANN)-based transformers for wireless foundation models (WFMs). Inspired by the noise-robust and energy-efficient information processing in the human brain, SpikeWFM aims to enhance the resilience of WFMs against noise and interference while maintaining strong generalization capabilities across diverse wireless scenarios.

Drawing from the success of large language models, WFMs leverage self-supervised pre-training on large-scale datasets spanning various wireless environments to learn a unified embedding that supports a wide range of downstream tasks, including channel prediction, channel estimation, beam predition, positioning and etc. Such models typically outperform task-specific designs and exhibit superior adaptability to unseen conditions. However, existing WFMs remain vulnerable to realistic noise and interference in practical wireless systems.

To address this limitation, we incorporate spiking neurons into the transformer-based WFM architecture. We provide a brief theoretical analysis demonstrating how the SNN-ANN hybrid effectively mitigates noise and interference through temporal sparsity and event-driven processing. Experimental results show that SpikeWFM consistently outperforms conventional ANN-based WFMs in both pre-training convergence and channel prediction accuracy. Additional results on communication and sensing tasks will be presented in the full journal version of this work.
\end{abstract}


\section{Introduction and Related Work}

The vision of sixth-generation (6G) wireless systems centers on the profound convergence of communication, sensing, and pervasive intelligence. This integration is poised to enable groundbreaking applications, including low-altitude aerial networks, embodied intelligent agents, and autonomous Vehicle-to-Everything (V2X) ecosystems. As 3GPP advances toward an AI-native air interface in Releases 18–20, there is an urgent need for scalable neural architectures that overcome the constraints of traditional task-specific deep learning models. While such models have shown success in isolated tasks like channel prediction, beam prediction, CSI compression, etc.~\cite{wang2025pilot, wang2025ps}, they are typically hand-engineered for fixed configurations and struggle to generalize across the highly dynamic, non-stationary, and interference-rich environments of real-world 6G deployments.
To address these scalability and generalization bottlenecks, the paradigm of wireless foundation models, also referred to as Big AI Models for wireless networks, has gained significant traction \cite{chen2024big}. 
By conducting large-scale self-supervised pre-training on massive and diverse wireless datasets, WFMs learn universal embeddings that can support a wide array of downstream tasks, such as channel prediction, beam management, sensing, and ISAC, through lightweight fine-tuning or even zero-shot inference \cite{chen2024big}.
Existing approaches in this domain generally follow two complementary technical trajectories. The first leverages cross-modal transfer from pre-trained large language models (LLMs) or vision models. Prominent examples include ChannelGPT, which embeds environment intelligence to generate digital twin channels for 6G pervasive intelligence \cite{yu2025channelgpt}, LVM4CSI that directly applies frozen large vision models to CSI-related tasks without task-specific fine-tuning \cite{guo2025lvm4csi}, as well as MUSE-FM for multi-task environment-aware modeling \cite{zheng2025muse}, LLM-based beam prediction \cite{sheng2025beam}, and other adaptations such as LLM4WM \cite{liu2025llm4wm}. The second trajectory emphasizes native wireless pre-training on raw physical-layer data, exemplified by WirelessGPT \cite{yang2025wirelessgpt2}, WiFo \cite{liu2025wifo}, WirelessJEPA \cite{chu2026wirelessjepa}, WiCC-Net\cite{jing2026signal} and BERT-inspired architectures for massive MIMO CSI prediction \cite{catak2025bert4mimo}. Recent discussions further underscore the evolution of LLMs in wireless systems from mere adaptation toward higher levels of autonomy \cite{liang2026large}.

Notwithstanding their strong zero-shot and few-shot capabilities, the majority of current WFMs are built upon standard artificial neural networks with continuous activation functions. In practical 6G scenarios characterized by hardware impairments, high-power interference, and low signal-to-noise ratio (SNR) conditions, these models frequently experience notable performance degradation. Continuous activations (e.g., ReLU) tend to introduce a non-zero noise floor, resulting in representation drift and cumulative DC bias across deep layers. Additionally, the substantial computational and energy demands of Transformer-based backbones pose significant challenges for deployment on energy-constrained edge devices and mobile terminals.

Spiking Neural Networks present a compelling bio-inspired alternative by mimicking the event-driven, sparse, and energy-efficient processing mechanisms of the human brain. Through discrete spatio-temporal spike encoding, SNNs inherently function as a “noise gate,” whereby the desired signal integrates linearly over time while stochastic noise and interference accumulate sub-linearly, thereby offering robust interference suppression \cite{auge2021survey, liu2024energy}. An early hybrid effort, SWS-Net \cite{lu2026spiking}, demonstrated the potential of lightweight ANN-SNN architectures for WiFi-based indoor sensing; however, it remains confined to task-specific applications within localized domains.

In this paper, we make a significant step forward by proposing SpikeWFM, a hybrid spiking-aided wireless foundation model. Distinct from prior task-specific hybrids, SpikeWFM integrates Leaky Integrate-and-Fire (LIF) neurons directly into a large-scale Transformer backbone. This design extends the localized noise resilience and temporal sparsity of SNNs into a scalable, generalizable representation learning framework suitable for diverse and noisy wireless environments. To the best of our knowledge, SpikeWFM represents the first SNN-Transformer hybrid developed as a universal wireless foundation model.
The main contributions of this work are fourfold:
(1) Novel Hybrid Architecture: We propose SpikeWFM, the first wireless foundation model that integrates spiking neurons into a Transformer backbone, combining high-capacity self-attention with bio-inspired noise resilience.
(2) Theoretical Robustness and Generalization: We provide a theoretical analysis demonstrating that spiking neurons achieve linear SNR gains and act as a structural Information Bottleneck (IB), which enhances generalization by constraining representation entropy and suppressing environment-specific noise.
(3) Superior Performance and Generalization: Experiments show that SpikeWFM outperforms Transformers in convergence speed and channel prediction accuracy, especially exhibiting strong zero-shot generalization in unseen environments.

\section{System Model and Preliminary}

We consider a wideband MIMO-OFDM uplink communication scenario, where a Base Station (BS) equipped with a large-scale antenna array of $N_r$ elements serves a user equipment (UE) featuring $N_t$ antennas. The wireless channel is characterized by its frequency-selective and time-varying nature. Specifically, the received signal at the BS in the frequency domain, corresponding to the $k$-th subcarrier and the $m$-th OFDM symbol, is expressed as $\mathbf{Y}_{m,k} = \mathbf{H}_{m,k}\mathbf{X}_{m,k} + \mathbf{N}_{m,k}$. Here, $\mathbf{H}_{m,k} \in \mathbb{C}^{N_r \times N_t}$ denotes the complex-valued frequency-response matrix. 

In practical deployments, due to the inherent constraints of pilot-assisted channel estimation, such as limited pilot overhead, hardware impairments, and non-ideal synchronization, the base station (BS) cannot obtain perfect channel state information (CSI). Instead, it acquires only a coarse and imperfect channel estimate: 
\begin{equation}
	\tilde{\mathbf{H}}_{m,k} = \mathbf{H}_{m,k} + \mathbf{E}_{m,k}
\end{equation}
where $\mathbf{E}_{m,k}$ represents the aggregate estimation error, which encapsulates additive white Gaussian noise (AWGN) at the receiver and residual multi-user interference from adjacent cells. 

To facilitate effective processing by high-capacity foundation models, the complete CSI tensor $\tilde{\mathcal{H}} \in \mathbb{C}^{M \times K \times N_r}$ is first organized and then partitioned into $N$ distinct spatio-temporal patches. Each patch, representing a local slice of the channel's delay-doppler-angular profile, is flattened and projected into a latent space via a trainable linear operator $\mathbf{W}_{emb} \in \mathbb{R}^{d \times L_{elem}}$. Following this high-dimensional embedding stage, the latent representation of a single CSI token can be abstractly modeled as a composite signal within the $d$-dimensional feature space:
\begin{equation}
	x = s + n, \quad s \in \mathbb{R}^d, n \sim \mathcal{N}(0, \sigma_n^2 \mathbf{I})
\end{equation}
In this formulation, $s$ signifies the invariant channel semantics, such as the underlying multipath geometry and stable scattering clusters, while $n$ represents the embedding-space noise induced by the raw estimation error $\mathbf{E}_{m,k}$. The primary objective of the proposed Spike-WFM framework is to design an operator that extracts a robust feature $f(x)$, maximizing the mutual information with the semantic component $s$ while effectively suppressing the stochastic and detrimental perturbation $n$.

\section{Robustness and Generalization Analysis}

Conventional wireless foundation models predominantly utilize continuous Feed-Forward Networks (FFNs) with Rectified Linear Unit (ReLU) activations. We now provide a theoretical demonstration of why Spike-WFM, powered by Leaky LIF neurons, offers inherently superior noise resilience and cross-environment generalization compared to its continuous counterparts.

\subsection{The DC Bias Problem in Continuous FFNs}

In standard Transformer-based architectures, the hidden activation for the $i$-th dimension is typically computed as $y_i = \max(0, w_i^T x + b_i)$. In low-SNR environments, the meaningful signal component $\mu_{z,i} = w_i^T s$ is frequently overwhelmed by the noise component $z_{n,i} = w_i^T n$, where $z_{n,i} \sim \mathcal{N}(0, \sigma_W^2)$ assuming normalized weights. When the input becomes noise-dominated (i.e., $\mu_{z,i} \to 0$), the expected output of the neuron does not vanish. Instead, the mathematical expectation of the activation is the mean of a rectified Gaussian distribution:
\begin{equation}
	\mathbb{E}[y_i] \approx \int_{0}^{\infty} z \cdot \frac{1}{\sqrt{2\pi}\sigma_W} e^{-\frac{z^2}{2\sigma_W^2}} dz = \frac{\sigma_W}{\sqrt{2\pi}}
\end{equation}
This non-zero value $\frac{\sigma_W}{\sqrt{2\pi}}$ constitutes a systemic Direct Current (DC) bias. In deep hierarchical architectures, this bias does not remain isolated; it accumulates across $L$ successive layers, resulting in "representation drift." This phenomenon occurs when the cumulative noise floor gradually saturates the dynamic range of the neurons, eventually leading to representation collapse where the model loses its discriminative power. Consequently, ReLU-based WFMs often struggle to distinguish subtle, high-frequency scattering components from the pervasive environmental noise.

\subsection{LIF-driven Spatio-Temporal Noise Gating}

Spike-WFM effectively addresses the DC bias issue by substituting ReLU with LIF neurons, which integrate feature information over $T$ discrete simulation steps. The membrane potential $V_i(t)$ of the neuron evolves according to the following temporal dynamics:
\begin{equation}
	V_i(t) = \tau V_i(t-1) + z_i(t) - S_i(t-1)V_{th}
\end{equation}
where $\tau \in (0, 1]$ represents the leak factor and $V_{th}$ denotes the firing threshold. By interpreting the integration process as a coherent temporal accumulation of the same CSI token, the signal component $\sum \mu_{z,i}$ grows linearly with the time window $T$. In contrast, the stochastic noise component $z_{n,i}(t)$ behaves as a random walk. For a leak factor $\tau \approx 1$, the internal SNR after $T$ steps, $\text{SNR}_{int}$, can be derived as:
\begin{equation}
	\text{SNR}_{int} = \frac{(\mathbb{E}[\sum_{t=1}^T z_i(t)])^2}{\text{Var}(\sum_{t=1}^T z_i(t))} = \frac{(T\mu_{z,i})^2}{T\sigma_W^2} = T \cdot \text{SNR}_{in}
\end{equation}
This linear gain in SNR (by a factor of $T$) enables the neuron to accumulate evidence of weak channel features that would otherwise be lost. Crucially, by carefully calibrating $V_{th} > \mathbb{E}[V_i(T)|s=0]$, the LIF neuron functions as a non-linear spatio-temporal gate. Sub-threshold noise fluctuations are naturally dissipated by the leak factor $\tau$ over time, effectively neutralizing the systemic DC bias. Only those features exhibiting sufficient spatio-temporal consistency can trigger a discrete spike $S_i(t)=1$, yielding a sparse and inherently denoised representation.

\subsection{Generalization via Structural Information Bottleneck}

The generation of binary spike trains $S \in \{0, 1\}^D$ imposes a structural Information Bottleneck (IB) that significantly enhances the model's generalization to unseen wireless environments. According to the IB principle, an optimal model seeks to compress the input representation by minimizing the mutual information $I(X; S)$ while retaining predictive power by maximizing $I(S; Y)$. The discrete and sparse nature of spikes imposes a hard limit on the channel capacity of the representation:
\begin{equation}
	I(X; S) \le \mathcal{H}(S) \le \sum_{i=1}^{D} \mathcal{H}_b(p_i)
\end{equation}
where $\mathcal{H}_b(p_i) = -p_i \log p_i - (1-p_i)\log(1-p_i)$ is the binary entropy associated with the firing probability $p_i$. In Spike-WFM, the gating effect drives $p_i$ toward 0 for dimensions dominated by noise, while robust semantic features drive $p_i$ toward 1. This resulting bi-modal distribution of firing probabilities significantly lowers the total representation entropy $\mathcal{H}(S)$ compared to continuous, high-variance activations. 

Applying the PAC-Bayesian framework for generalization, the gap between training and testing error $\Delta \mathcal{R}$ is bounded by $\mathcal{O}(\sqrt{I(X; S)/N_{train}})$. By structurally constraining the flow of information through sparse spiking patterns, Spike-WFM prevents the model from "memorizing" environment-specific noise or site-specific multipath artifacts (overfitting). This ensures superior zero-shot performance when the model is deployed across diverse and previously unobserved wireless channel realizations.

\section{Proposed Spike-WFM Architecture}

\begin{figure}[htbp]
	\centering
	\includegraphics[width=1\linewidth]{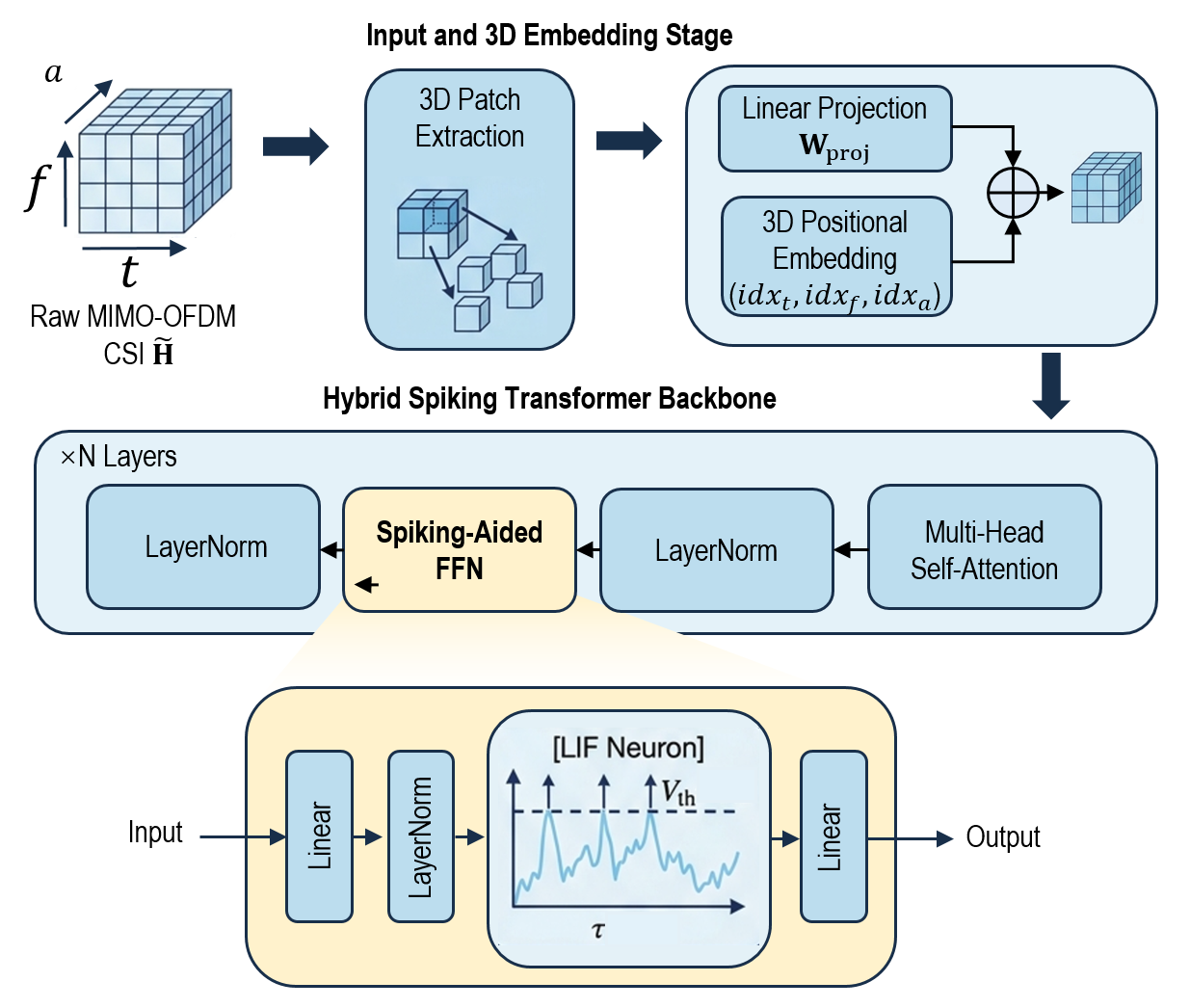}
	\caption{SpikeWFM Model Architecture: From 3D CSI partitioning to spiking latent representation and downstream reconstruction.}
	\label{fig:spike-wfm-arch}
\end{figure}

The Spike-WFM architecture is specifically engineered to extract robust channel representations from high-dimensional, noisy CSI tensors. As illustrated in the detailed schematic in Fig. \ref{fig:spike-wfm-arch}, the model pipeline is organized into three core functional modules: a 3D spatio-temporal-frequency embedding layer, a hybrid spiking transformer backbone, and a task-specific prediction head.

\subsection{Spatio-Temporal Tokenization and 3D Embedding}

To fully preserve the intricate multi-dimensional correlations inherent in MIMO-OFDM signals, the input CSI matrix $\tilde{\mathbf{H}}$ is first partitioned into non-overlapping 3D patches that span the time, frequency, and antenna dimensions. Each 3D patch is flattened into a compact vector $\mathbf{t}_n \in \mathbb{R}^{L_{elem}}$, where $L_{elem}$ is set to 336 in our configuration to balance granularity and complexity. These vectors are then projected into a high-dimensional $d_{model}$-dimensional embedding space ($d_{model} = 640$) via a trainable linear layer $\mathbf{W}_{proj}$.

To maintain the global coordinate context within the 3D CSI grid, we utilize a decoupled 3D positional embedding mechanism. Recognizing that the physical properties of the time, frequency, and antenna domains differ significantly, our model learns three independent sets of embeddings. These are concatenated along the feature dimension to form the final encoding:
\begin{equation}
	\mathbf{E}_{pos} = [\text{Emb}_t(idx_t) \parallel \text{Emb}_f(idx_f) \parallel \text{Emb}_a(idx_a)]
\end{equation}
where $\parallel$ denotes the concatenation operation, with sub-dimensions allocated to sum up to $d_{model}$. The indices $idx_t, idx_f, idx_a$ correspond to the patch's coordinates in the respective domains. The resulting input to the transformer backbone is then normalized: $\mathbf{x}_0 = \text{LayerNorm}(\mathbf{W}_{proj}\mathbf{t}_n + \mathbf{E}_{pos})$.

\subsection{Hybrid Spiking Foundation Backbone}

The backbone of Spike-WFM consists of $N=16$ stacked hybrid encoder layers, designed for deep feature extraction. Each layer combines the global dependency modeling of Multi-Head Self-Attention (MHSA) with the noise-resilient properties of a Spiking Feed-Forward Network (SFFN).

The SFFN module is the critical component where conventional ReLU is replaced by LIF neurons to introduce bio-inspired temporal integration. For a given hidden state $\mathbf{h}$, the transformation within the foundation backbone is defined as follows:
\begin{equation}
	\mathbf{z} = \text{LayerNorm}(\mathbf{W}_{fc1} \mathbf{h})
\end{equation}
\begin{equation}
	V(t) = \tau V(t-1) + \mathbf{z}(t), \quad \mathbf{s}(t) = \mathcal{H}(V(t) - V_{th})
\end{equation}
In this process, $V(t)$ tracks the membrane potential, $\tau = 5.0$ acts as the decay factor to regulate information leakage, and $\mathcal{H}(\cdot)$ is the Heaviside step function that governs spike generation. The resulting spike output $\mathbf{s}(t)$ is subsequently projected back to the original model dimension through $\mathbf{W}_{fc2}$. This spiking mechanism functions as a rigorous filter, effectively blocking low-amplitude noise components that fail to reach the threshold $V_{th}=1.5$, thereby ensuring a sparse and robust latent representation.

\subsection{Downstream Channel Prediction Head}

To adapt the pre-trained foundation backbone for the specific task of channel prediction, we append a specialized transformer-based head. The latent features emerging from the backbone are first projected into a dedicated prediction space with dimension $d_{pred} = 512$. We also introduce 1D sinusoidal positional encodings at this stage to further emphasize the sequential dependencies required for accurate forecasting or interpolation.

The prediction head is composed of $K=3$ Custom Spiking Encoder layers. While these layers maintain the MHSA-SFFN structural paradigm, they are specifically optimized for sequence-to-sequence regression. Notably, to ensure stable temporal dynamics and consistent feature distributions during the fine-tuning phase, the SFFN within the prediction head utilizes 1D Batch Normalization instead of Layer Normalization prior to the LIF integration. 

Finally, the processed tokens are passed to a linear decoder for CSI reconstruction:
\begin{equation}
	\hat{\mathbf{H}}_{pred} = \mathbf{W}_{dec} \mathbf{x}_{final} + \mathbf{b}_{dec}
\end{equation}
where $\mathbf{W}_{dec}$ maps the $d_{pred}$ hidden features back to the physical CSI dimension ($L_{elem}=336$). By building upon the noise-suppression capabilities of the backbone, the prediction head can reconstruct the target CSI with high fidelity, even in the presence of severe interference.

\section{Experiments}

In this section, we evaluate the proposed SpikeWFM through a four-part experimental framework. First, we establish a high-fidelity simulation environment using the DeepMIMO dataset with a robust scenario-wise data allocation. Second, we analyze the foundation backbone’s convergence and noise-filtering efficacy during self-supervised pre-training. Finally, we benchmark both in-domain and cross-domain prediction performance against traditional and ANN-based methods across varying SNR levels.

\subsection{Experiments Setups}

We conduct extensive simulations using the DeepMIMO dataset to evaluate the performance of the proposed SpikeWFM. The system considers a wideband MIMO-OFDM uplink scenario operating at a carrier frequency of 3.5 GHz. The base station (BS) and user equipment (UE) are both equipped with a 2- Uniform Linear Arrays (ULAs) with a half-wavelength spacing. To capture intricate channel dynamics, we utilize 612 subcarriers with a spacing of 30 kHz. The temporal dimension is configured with $n=4$ segments, resulting in $14 \times 4$ OFDM symbols per realization. The user speed is randomly sampled within the range of 0.1 to 5.0 m/s, and each channel is generated considering 25 distinct propagation paths to reflect realistic multipath scattering environments.

The hybrid spiking foundation backbone is pre-trained using CSI data aggregated from seven diverse city scenarios. To maintain balanced environmental coverage, we implement a scenario-wise sample allocation, where each city scenario contributes 1,500 samples for training and 500 for validation, resulting in a comprehensive pre-training dataset of 14,000 samples in total. Raw CSI tensors are partitioned into non-overlapping 3D patches and projected into a 640-dimensional embedding space.To cater to diverse application scenarios with varying latency and performance requirements, we have developed three versions of the foundation model with 3.5M, 20M, and 80M parameters. In this paper, we select the 80M parameter version as the primary backbone to rigorously evaluate the effectiveness of the Spike-WFM architecture. We employ the Adam optimizer with an initial learning rate of $5 \times 10^{-5}$ and a StepLR scheduler. The backbone is trained for 200 epochs with an early stopping patience of 20.


For the downstream channel prediction tasks (both In-Domain and Cross-Domain), the pre-trained backbone is frozen as a feature extractor. For In-Domain tasks, the training set comprises 22,800 samples aggregated from 7 cities, with 5,700 samples each for validation and testing; for Cross-Domain tasks, the training set also contains 22,800 samples from 7 cities, while validation and testing sets consist of 4,800 samples each from 3 unseen cities, ensuring all scenarios consistently adhere to a 12:3:3 allocation ratio. The prediction head, consisting of 3 spiking encoder layers, is optimized using NMSE Loss and the Adam optimizer (learning rate: $3 \times 10^{-4}$) for a maximum of 50 epochs with an early stopping patience of 5, emphasizing the efficiency of the lightweight downstream model.

\subsection{Analysis of Spiking-Aided Foundation Backbone}
\begin{figure}[htbp]
	\centering
	\includegraphics[width=1\linewidth]{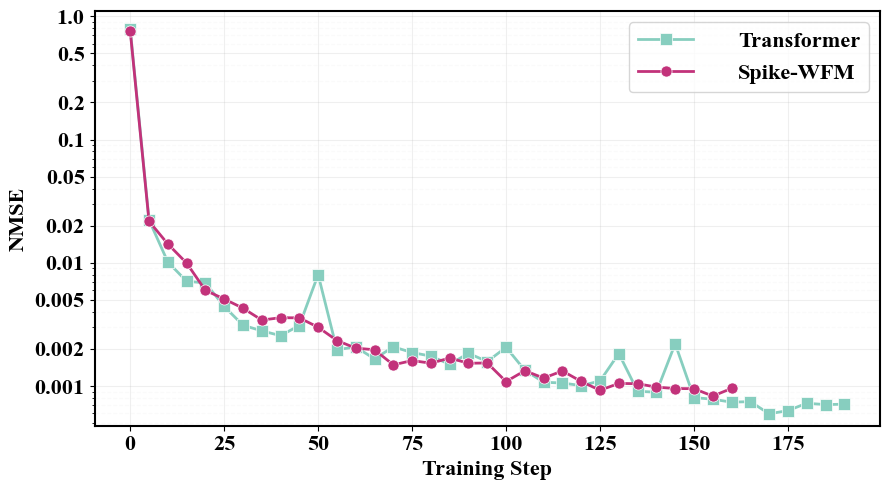}
	\caption{Comparison of pre-training validation loss (NMSE). We compare the proposed SpikeWFM against a conventional ReLU-based Transformer baseline using a self-supervised masked autoencoding (MAE) strategy (mask ratio = 0.75). The SpikeWFM integrates spiking mechanisms via LIF neurons ($\tau=5.0$, $V_{th}=1.0$, ATan surrogate gradient) to reconstruct the complete CSI tensor from sparse observations.}
	\label{fig:fundation_loss}
\end{figure}

As illustrated in Fig. \ref{fig:fundation_loss}, we evaluate the representation learning capability of SpikeWFM through its convergence behavior. Both the spiking-aided backbone and the Transformer exhibit a rapid decrease in reconstruction error during the initial stages.

Notably, SpikeWFM demonstrates superior convergence efficiency, reaching a stable performance plateau significantly faster than the ANN-based baseline. This rapid adaptation suggests that the discrete spiking mechanism effectively identifies and prioritizes dominant multipath features in the CSI data.

From a physiological perspective, the LIF neurons in SpikeWFM function as a temporal "noise gate." By integrating consistent signal components while dissipating stochastic perturbations through the leak factor, the model triggers spikes only when the membrane potential exceeds the threshold. This mechanism filters out low-amplitude interference, yielding a sparse and robust latent representation early in the pre-training phase. Although the ReLU-based Transformer eventually reaches a marginally lower validation loss, its trajectory is characterized by more pronounced fluctuations and delayed convergence compared to the stability offered by the spiking neurons.
\subsection{In-Domain Prediction Performance}

\begin{figure}[htbp]
	\centering
	\includegraphics[width=1\linewidth]{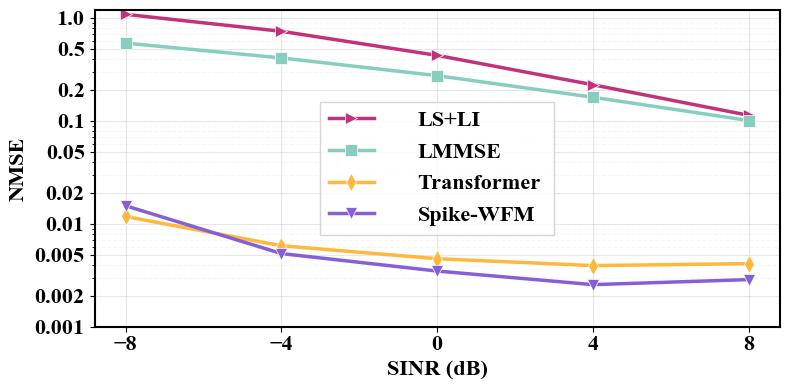}
	\caption{In-domain channel prediction performance under varying noise intensities. We evaluate the NMSE of SpikeWFM across SINR levels from $-8$ dB to $8$ dB. During downstream training, the model is exposed to random noise within this range to enhance robust representation learning, while its performance is validated under fixed noise levels. Note that SpikeWFM performs prediction without requiring explicit SNR values as input.}
	\label{fig:indomain}
\end{figure}

We evaluate the channel prediction accuracy of SpikeWFM across a range of SINR levels from $-8$ dB to $8$ dB. As illustrated in Fig. \ref{fig:indomain}, SpikeWFM consistently maintains a leading edge in prediction accuracy compared to traditional signal processing and deep learning benchmarks. A key observation is that the model achieves its optimal performance at approximately $4$ dB, where the spiking mechanism effectively balances noise suppression and feature preservation.However, two specific boundary phenomena merit further technical discussion. At the extreme low SNR of $-8$ dB, the vanilla Transformer baseline exhibits a marginal advantage over SpikeWFM. This can be attributed to the "purity bias" of our foundation backbone, which was pre-trained solely on a masking objective without noise intervention, thus optimizing the latent space for structural completion rather than extreme denoising. In such high-interference scenarios, the discrete firing threshold of LIF neurons may inadvertently suppress faint but critical signal components that the Transformer’s continuous activations can still partially recover.As the SINR improves beyond $-4$ dB, SpikeWFM quickly surpasses all baselines, demonstrating a much flatter and more stable performance slope. Interestingly, both AI-based models show a slight loss increase at $8$ dB compared to $4$ dB. This suggests a subtle distribution shift; since the downstream model is trained with stochastic noise injection between $-8$ dB and $8$ dB, the "too clean" environment at $8$ dB deviates from the noise-inclusive distribution learned during fine-tuning. 

\subsection{Cross-Domain Generalization Analysis}

\begin{figure}[htbp]
	\centering
	\includegraphics[width=1\linewidth]{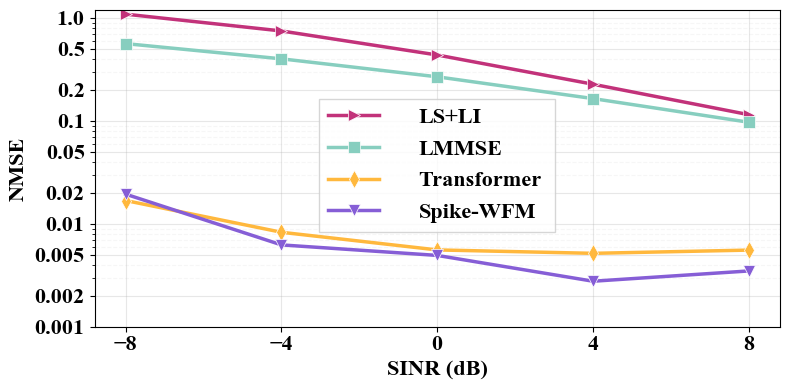}
	\caption{Cross-domain channel prediction performance across unseen environments. We evaluate the generalization capability of SpikeWFM on three independent city scenarios that were entirely excluded from both pre-training and downstream training phases. The NMSE is measured across SINR levels from $-8$ dB to $8$ dB. During fine-tuning, the model is trained on seven different cities with stochastic noise injection to learn environment-agnostic representations, while testing is conducted under fixed noise levels without explicit SNR priors.}
	\label{fig:crossdomain}
\end{figure}
To evaluate the universal applicability of the proposed foundation model, we conduct a cross-domain generalization analysis by testing on three independent city scenarios that were entirely excluded from both the pre-training and downstream training phases. Specifically, the model is trained on data from seven diverse cities with stochastic noise injection ranging from $-8$ dB to $8$ dB, and then evaluated on the remaining three unseen cities to assess its zero-shot transferability across different geographical multipath profiles. As illustrated in Fig. \ref{fig:crossdomain}, SpikeWFM maintains a consistent performance advantage in the majority of SINR conditions, proving that the learned spiking representations are effectively environment-agnostic.In the extreme low-SNR regime ($-8$ dB), the vanilla Transformer baseline exhibits a marginal lead, which reinforces our observation of a "purity bias" inherited from the foundation backbone pre-trained solely on masking objectives. In such scenarios, the discrete firing threshold of LIF neurons may suppress faint signal components heavily corrupted by noise in an unfamiliar domain. However, as the SINR improves beyond $-4$ dB, SpikeWFM quickly surpasses the Transformer-based baseline, demonstrating a flatter and more stable NMSE trajectory. Similar to the in-domain results, both AI-native models show slight performance saturation at $8$ dB due to the distribution shift from the noisy training environment to the nearly ideal test conditions.

\section{Conclusion and Future Work}
In this paper, we proposed SpikeWFM, a novel spiking-aided foundation model for robust channel prediction in future 6G networks. By integrating LIF neurons into a hybrid SNN-ANN Transformer architecture, we successfully addressed the challenges of systemic DC bias and representation drift inherent in conventional continuous backbones. Our experimental results demonstrate that the proposed spiking-aided temporal integration effectively functions as a noise gate, enabling the model to extract denoised and discriminative channel semantics even in extreme low-SNR regimes. Extensive evaluations on the DeepMIMO dataset show that SpikeWFM significantly outperforms traditional signal processing benchmarks and ANN-based models in both in-domain accuracy and cross-domain generalization, particularly highlighting its capability to capture environment-agnostic features across diverse city scenarios.

Future work will focus on expanding the utility of the pre-trained spiking foundation backbone to a broader range of downstream tasks. Specifically, we aim to extend SpikeWFM to support high-precision channel estimation and wireless localization, leveraging the SNN’s superior spatio-temporal resolution to resolve fine-grained multipath components in complex environments. Furthermore, we intend to investigate more robust pre-training paradigms to further bridge the performance gap in extreme edge-case scenarios and enhance the model’s adaptability across the full dynamic range of interference levels, without compromising its independence from explicit SNR priors.

\bibliographystyle{IEEEtran}
\bibliography{IEEEexample}

@article{liu2025wifo,
	title={WiFo-CF: Wireless Foundation Model for CSI Feedback},
	author={Liu, Xuanyu and Gao, Shijian and Liu, Boxun and Cheng, Xiang and Yang, Liuqing},
	journal={arXiv preprint arXiv:2508.04068},
	year={2025}
}

@article{jing2026signal,
  title={Signal Compression for Wireless Communication and Sensing: A General Approach Utilizing Pretrained Wireless Foundation Models},
  author={Jing, Liwen and Yang, Tingting and Zhang, Han and Shi, Yuxuan and Zhang, Chi and Zhang, Bowen},
  journal={IEEE Transactions on Mobile Computing},
  year={2026},
  publisher={IEEE}
}

@article{auge2021survey,
  title={A survey of encoding techniques for signal processing in spiking neural networks},
  author={Auge, Daniel and Hille, Julian and Mueller, Etienne and Knoll, Alois},
  journal={Neural Processing Letters},
  volume={53},
  number={6},
  pages={4693--4710},
  year={2021},
  publisher={Springer}
}

@article{liu2024energy,
  title={Energy-efficient distributed spiking neural network for wireless edge intelligence},
  author={Liu, Yanzhen and Qin, Zhijin and Li, Geoffrey Ye},
  journal={IEEE Transactions on Wireless Communications},
  volume={23},
  number={9},
  pages={10683--10697},
  year={2024},
  publisher={IEEE}
}

@inproceedings{lu2026spiking,
  title={Spiking-Aided Neural Architecture for Efficient and Robust WiFi Sensing},
  author={Lu, Yisha and Jing, Liwen and Zheng, Jiangmao and Zhang, Bowen},
  booktitle={Proceedings of the AAAI Conference on Artificial Intelligence},
  volume={40},
  number={29},
  pages={24106--24114},
  year={2026}
}

@article{chen2024big,
  title={Big AI models for 6G wireless networks: Opportunities, challenges, and research directions},
  author={Chen, Zirui and Zhang, Zhaoyang and Yang, Zhaohui},
  journal={IEEE wireless communications},
  volume={31},
  number={5},
  pages={164--172},
  year={2024},
  publisher={IEEE}
}

@article{yu2025channelgpt,
  title={ChannelGPT: A large model toward real-world channel foundation model for 6G environment intelligence communication},
  author={Yu, Li and Shi, Lianzheng and Zhang, Jianhua and Zhang, Zhen and Zhang, Yuxiang and Liu, Guangyi},
  journal={IEEE Communications Magazine},
  volume={63},
  number={10},
  pages={68--74},
  year={2025},
  publisher={IEEE}
}

@article{guo2025lvm4csi,
  title={LVM4CSI: Enabling direct application of pre-trained large vision models for wireless channel tasks},
  author={Guo, Jiajia and Jiang, Peiwen and Wen, Chao-Kai and Jin, Shi and Zhang, Jun},
  journal={arXiv preprint arXiv:2507.05121},
  year={2025}
}

@article{sheng2025beam,
  title={Beam prediction based on large language models},
  author={Sheng, Yucheng and Huang, Kai and Liang, Le and Liu, Peng and Jin, Shi and Li, Geoffrey Ye},
  journal={IEEE Wireless Communications Letters},
  year={2025},
  publisher={IEEE}
}

@article{yang2025wirelessgpt2,
  title={WirelessGPT: A Generative Foundation Model for Multi-Task Integrated Sensing and Communication},
  author={Yang, Tingting and Zhang, Ping and Zheng, Mengfan and Shi, Yuxuan and Jing, Liwen and Huang, Jianbo and Li, Nan},
  journal={IEEE Journal on Selected Areas in Communications},
  year={2025},
  publisher={IEEE}
}

@article{catak2025bert4mimo,
  title={BERT4MIMO: A foundation model using BERT architecture for massive MIMO channel state information prediction},
  author={Catak, Ferhat Ozgur and Kuzlu, Murat and Cali, Umit},
  journal={arXiv preprint arXiv:2501.01802},
  year={2025}
}

@article{liu2025llm4wm,
  title={LLM4WM: Adapting LLM for wireless multi-tasking},
  author={Liu, Xuanyu and Gao, Shijian and Liu, Boxun and Cheng, Xiang and Yang, Liuqing},
  journal={IEEE Transactions on Machine Learning in Communications and Networking},
  year={2025},
  publisher={IEEE}
}

@article{liang2026large,
  title={Large language models for wireless communications: From adaptation to autonomy},
  author={Liang, Le and Ye, Hao and Sheng, Yucheng and Wang, Ouya and Wang, Jiacheng and Jin, Shi and Li, Geoffrey Ye},
  journal={IEEE Communications Magazine},
  year={2026},
  publisher={IEEE}
}

@article{chu2026wirelessjepa,
	title={WirelessJEPA: A Multi-Antenna Foundation Model using Spatio-temporal Wireless Latent Predictions},
	author={Chu, Viet and Mashaal, Omar and Abou-Zeid, Hatem},
	journal={arXiv preprint arXiv:2601.20190},
	year={2026}
}

@article{zheng2025muse,
	title={Muse-fm: Multi-task environment-aware foundation model for wireless communications},
	author={Zheng, Tianyue and Guo, Jiajia and Dai, Linglong and Jin, Shi and Zhang, Jun},
	journal={arXiv preprint arXiv:2509.01967},
	year={2025}
}

@article{wang2025pilot,
	title={Pilot-Free OFDM Transmission for Vehicular Communications With Asymmetric Constellation and Two-Stage Receiver},
	author={Wang, Yuwei and Sun, Li and Du, Qinghe and Elkashlan, Maged},
	journal={IEEE Transactions on Vehicular Technology},
	year={2025},
	publisher={IEEE}
}

@article{wang2025ps,
	title={PS-Net: Position-based precoding with sensing assistance for MIMO downlink transmission},
	author={Wang, Yuwei and Sun, Li and Du, Qinghe and Elkashlan, Maged},
	journal={IEEE Transactions on Communications},
	volume={73},
	number={8},
	pages={6410--6422},
	year={2025},
	publisher={IEEE}
}


\end{document}